\documentclass[reprint,amsmath,amssymb,aps,prb,superscriptaddress]{revtex4-1}

\usepackage{textcomp,pxfonts}
\usepackage{amsmath}
\usepackage{graphicx}
\usepackage{multirow}

\usepackage{color}
\begin{document}



\title{Phonon quarticity induced by changes in phonon-tracked hybridization during lattice expansion and its stabilization of rutile TiO$_2$}


\author{Tian Lan} 
\email[]{tianlan@caltech.edu}
\affiliation{\mbox{Department of Applied Physics and Materials Science, \nolinebreak California Institute of Technology, Pasadena, California 91125, USA}}
\author{C. W. Li}
\affiliation{Materials Science and Technology Division, Oak Ridge National Laboratory, Oak Ridge, Tennessee 37831, USA}
\author{O. Hellman}
\affiliation{\mbox{Department of Applied Physics and Materials Science, \nolinebreak California Institute of Technology, Pasadena, California 91125, USA}}
\author{D. S. Kim}
\affiliation{\mbox{Department of Applied Physics and Materials Science, \nolinebreak California Institute of Technology, Pasadena, California 91125, USA}}
\author{J. A. Mu\~{n}oz}
\affiliation{\mbox{Department of Applied Physics and Materials Science, \nolinebreak California Institute of Technology, Pasadena, California 91125, USA}}
\author{H. Smith}
\affiliation{\mbox{Department of Applied Physics and Materials Science, \nolinebreak California Institute of Technology, Pasadena, California 91125, USA}}
\author{D. L. Abernathy}
\affiliation{Quantum Condensed Matter Division, Oak Ridge National Laboratory, Oak Ridge, Tennessee 37831, USA}
\author{B. Fultz}
\affiliation{\mbox{Department of Applied Physics and Materials Science, \nolinebreak California Institute of Technology, Pasadena, California 91125, USA}}



\date{\today}

\begin{abstract}

Although the rutile structure of TiO$_2$ is stable at high temperatures,
the conventional quasiharmonic approximation predicts that several  acoustic
phonons decrease anomalously to zero frequency with thermal expansion,
incorrectly predicting a structural collapse at temperatures well below 1000\,K.
Inelastic neutron scattering  was used to measure the 
temperature dependence of the phonon density of states (DOS) of rutile TiO$_2$ from 300 to 1373\,K.
Surprisingly, these anomalous acoustic phonons were found to increase in frequency with temperature.
First-principles calculations showed that with lattice  expansion, 
the potentials for the anomalous acoustic phonons transform 
from  quadratic  to  quartic, stabilizing the rutile phase at high temperatures. 
In these  modes, the vibrational displacements of adjacent Ti and O atoms 
cause variations in hybridization of  $3d$  electrons of Ti and $2p$ electrons of O atoms.
With thermal expansion, the energy variation in this ``phonon-tracked hybridization" 
flattens the bottom of the interatomic potential well between Ti and O atoms, and induces a quarticity
in the phonon potential. 

\end{abstract}


\maketitle


\section{Introduction}

Titanium dioxide (TiO$_2$) is of longstanding interest in physics, chemistry, surface science and materials science. It has 
numerous technological applications  including pigments, optical coatings, catalysis, solar cells and gas sensors.\cite{ChemRev1995, SurfRev2003, Du2009, Lu2009}
Rutile is the stable phase of TiO$_2$ at  temperatures
below 1800\,K, and
the two other naturally-occurring phases
of TiO$_2$, anatase and brookite, both convert to rutile upon heating.\cite{Isaak1998}
The thermal properties of rutile TiO$_2$ are central for applications  that involve heat generation, thermal transport and temperature-driven phase transitions. 
Nevertheless, even after numerous experimental \cite{Smith2009, Zhang1998, Neutron1971, Samara1973, Burdett1987} and  theoretical  \cite{Lee1994, Vast2002,  Muscat2002,  Harrison2004, Labat2007, Mitev2010, Refson2013,Sikora2005, Lan2012} 
investigations, the  stability of the rutile phase at high temperatures remains poorly understood.

The quasiharmonic approximation (QHA) is based on how phonon frequencies change with volume. 
In the QHA, all shifts of phonon
frequencies from their low temperature values are the result of thermal expansion alone.\cite{Fultz2010, Chen2015} 
For a lattice expansion of 0.5\%, the energies of transverse phonons
decrease by 10\% to 50\% in the QHA, giving  Gr\"{u}neisen parameters as large as 100 for rutile TiO$_2$.\cite{Mitev2010, Refson2013}
In the QHA, many transverse acoustic modes decrease to zero frequency at a lattice expansion of 0.75\%, so the QHA
predicts a collapse of the rutile structure at temperatures  below 1000 \,K.
The QHA does not  account for 
all non-harmonic effects, however. 
Although the QHA accounts for some frequency shifts, 
the phonon modes are still assumed to be harmonic, non-interacting, 
and their energies depend only on the volume
of the crystal.

In non-harmonic potentials, phonon-phonon interactions are responsible for 
pure anharmonicity that shortens phonon lifetimes and shifts phonon frequencies. 
Anharmonicity competes with quasiharmonicity to alter the stability of phases at high temperatures, 
as has been shown, for example, with experiments and frozen phonon calculations 
on bcc Zr \cite{Ye1987} and
the possible stabilization of bcc Fe-Ni alloys at conditions of the Earth's core.\cite{Dubrovinsky2007}
For PbTe and ScF$_3$, there are recent reports of anharmonicity  being so large
that both the QHA and anharmonic perturbation theory 
fail dramatically.\cite{Delaire2011, ScF3}
These cases are suitable for ab initio
molecular dynamics (AIMD) simulations, however, which naturally 
includes all orders of anharmonicity.\cite{Koker2009, MD2010, Lan2014}
The AIMD method should be reliable when the electrons 
are near their ground states and the nuclear motions are classical. 

In this work, we study phonons at high temperatures
in rutile TiO$_2$
by  neutron inelastic scattering, AIMD, 
and  other methods. 
We report a remarkable quartic anharmonicity that develops 
because the hybridization between electron states at Ti and O 
atoms changes dynamically during atom vibrations, and these changes are
highly sensitive to lattice expansion. This ``phonon-tracked hybridization" 
induces the phonon quarticity that is essential for
stabilizing the rutile phase at high temperatures, and affects properties such as ferroelectricity and thermal transport. 


\begin{figure*}[t]
\includegraphics[width=1.8\columnwidth]{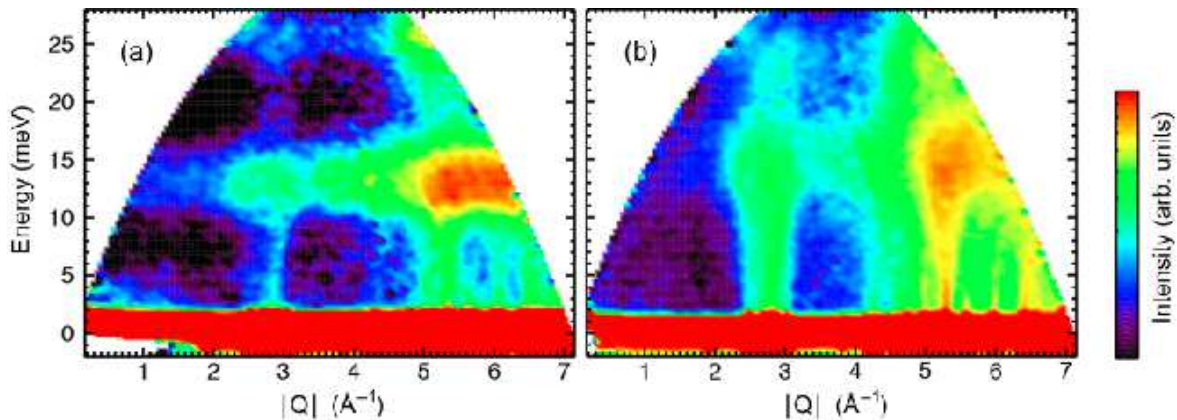}
\caption{
$S(Q,E)$ spectra of rutile TiO$_2$ measured at (a) 300K and (b) 1373K, respectively. The incident energy of neutron beam is 30 meV.
}
\label{fig:sqe}
\end{figure*}

\section{Experiment}

\subsection{Inelastic neutron scattering}

Samples were commercial TiO$_2$ powder (Alfa Aesar, Ward Hill, MA) with a rutile phase fraction of at least 99.9\%. 
The sample powder was loaded into a niobium sachet for neutron scattering measurements at high temperatures. 
Measurements were performed with the ARCS spectrometer at the Oak Ridge National Laboratory \cite{arcs},
using incident neutron energies of 75\,meV and 30\,meV. 
A vacuum furnace with niobium radiation shields was used for measurements at 300, 673, 1073, and 1373 K. 
Furnace backgrounds were measured with empty sample containers at each temperature.
Data reduction was performed with the software package DRCS.\cite{DRCS}
The raw data of individual neutron detection events were first binned into scattering angle and  energy transfer,  $E$,
and normalized by the proton current on target. 
The data were corrected for furnace background and  detector efficiency. 
The data were then rebinned into intensity, $S(Q,E)$, where $\hbar Q$ is the momentum transfer to the sample.  
This $S(Q,E)$ is averaged over all crystallographic directions, and is a reasonably well-balanced sampling of
all phonons in the material.

\begin{figure}[]
\includegraphics[width=0.85\columnwidth]{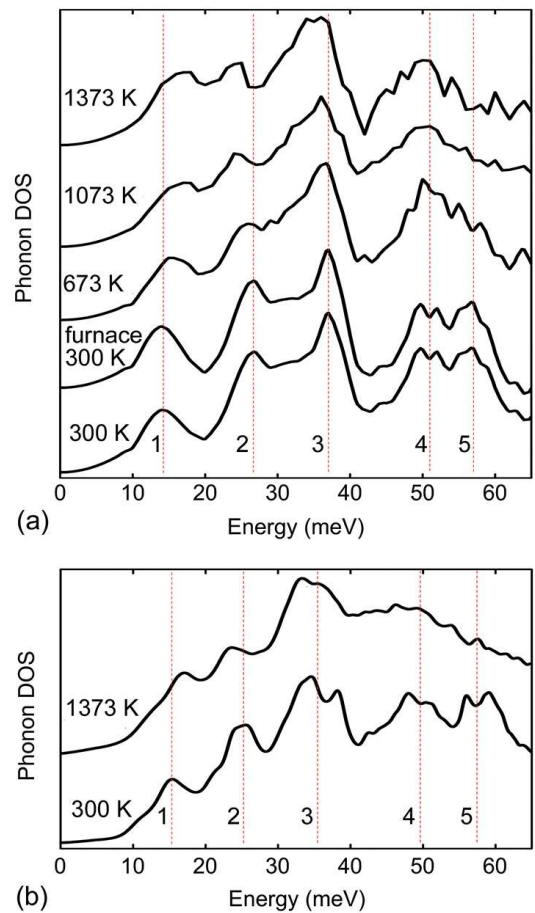}
\caption{(a) Neutron weighted phonon DOS of rutile TiO$_2$ from measurements at temperatures from 300 to 1373\,K with an incident energy of 75\,meV. Five vertical lines are aligned to peak centers at 300\,K.
(b) Phonon DOS of rutile TiO$_2$ at temperatures of 300 and 1373\,K calculated with first principles MD simulations and the FTVAC method.}
\label{fig:DOS}
\end{figure}

\begin{figure}
\includegraphics[width=0.76\columnwidth]{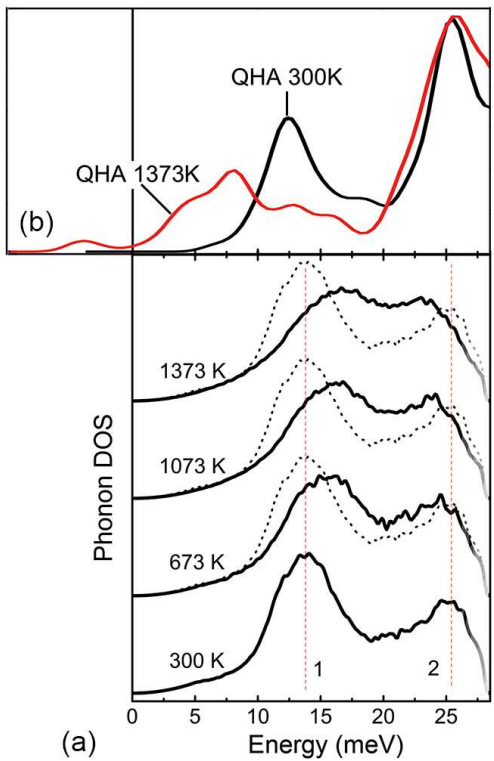}
\caption{(a) Neutron weighted phonon DOS of rutile TiO$_2$ from measurements at temperatures from 300 to 1373\,K with an incident energy of 30\,meV (so data above 26\,meV are less reliable). 
The dashed spectrum corresponds to the experimental result at 300\,K, shifted
vertically for comparison at each temperature.
(b) Simulated total phonon DOS from the QHA at 300\,K (black) and 1373\,K (red).}
\label{fig:dos30}
\end{figure}

\subsection{Results} 

Surprisingly, temperature causes
a global  increase in energy of the entire TA branch.
The lowest phonon dispersions formed by the TA branch at 14 meV at 300 K increase significantly in energy at 1373\,K,
as  seen in the $S(Q, E)$ phonon spectra (Fig.~\ref{fig:sqe}). 
There are no dispersions that soften between 2 and 4 $\AA^{-1}$, and there are 
no signs of phonons that collapse to zero frequency. 
This anomalous thermal stiffening  is also apparent in 
the phonon DOS curves, which were obtained after corrections for multiphonon and multiple scattering using the \emph{getdos} 
package.\cite{kresch08}
As shown in Fig. \ref{fig:DOS}(a), 
peak 1 (centered at 14\,meV) exhibits an unusual stiffening as large as 2.7\,meV,  
and it remains sharp.  
The large stiffening of this DOS peak is even more apparent in measurements
with higher energy resolution, using an incident energy of 30\,meV (see Fig.~\ref{fig:dos30}(a)).
On the other hand, as shown in Fig.~\ref{fig:DOS}(a) and Fig.~\ref{fig:dos30}(a), most other features of the phonon DOS above 20\,meV undergo substantial softening and broadening with temperature, in good agreement with the individual phonons measured by Raman spectrometry
at high temperatures.\cite{Samara1973, Lan2012}
It is well-known that rutile TiO$_2$ has a ferroelectric soft mode A$_{2u}$ at the $\Gamma$-point that stiffens  with heating.\cite{Samara1973} but the global stiffening of the entire lower-lying TA branch is surprising.

 


Figure \ref{fig:dos30}(b)
presents the DOS calculated with the quasiharmonic model at 1373\,K, showing 
that contrary to the experimental trend in Fig. \ref{fig:dos30}(a), 
the first peak in the phonon DOS undergoes
a large softening with temperature.
Consistent with previous results,\cite{Mitev2010, Refson2013}
our QHA calculations predict that modes in the TA branch soften
through zero frequency with the thermal expansion of
1373\,K, giving imaginary frequencies  in the DOS  (shown as negative in Fig. \ref{fig:dos30}(b)) 
that would destabilize the rutile structure at high temperatures. 

\begin{figure*}[t]
\includegraphics[width=2.15\columnwidth]{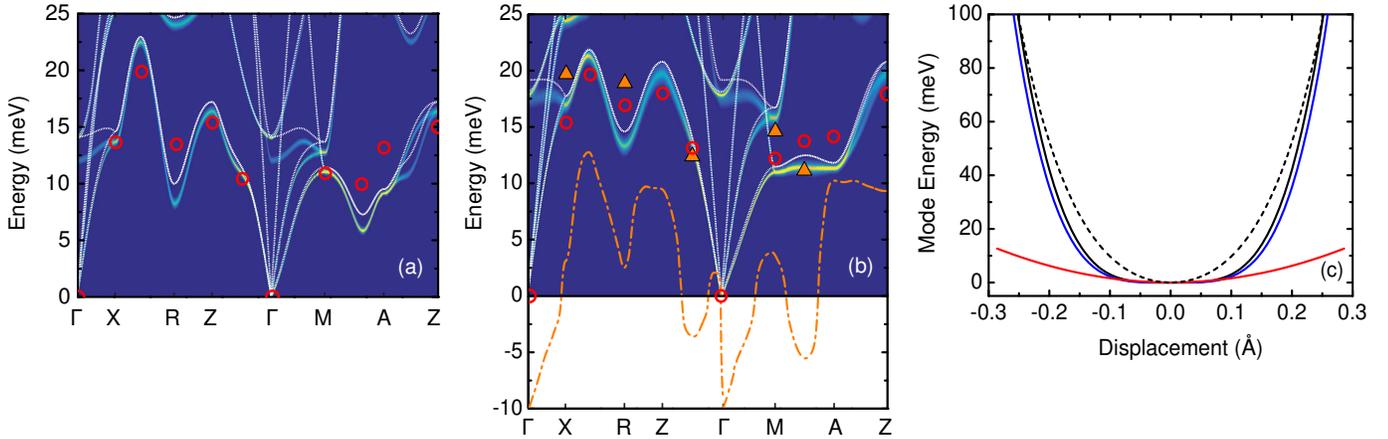}
\caption{Diffuse curves are 
TDEP phonon dispersions 
below 25\,meV at (a) 300\,K, and (b) 1373\,K,  compared with the results from the FTVAC method (red circles). The white curves are phonon dispersions for the quasiharmonicity plus quartic anharmonicity calculated with all $\psi_{ijk}$ set to zero in Eq.~\ref{eq:hamil}.  In (b), the dispersions are compared to the quasiharmonic dispersions (orange dashed curve) and the  single quartic oscillator model (orange triangles).  
(c) Frozen phonon potential (black) of TA mode  at R point with $q = (0.5, 0, 0.5)$ at 1373\,K,
showing the harmonic component (red) and quartic component (blue). The 
low temperature potential surface is also shown (dashed black). 
}
\label{fig:quartic}
\end{figure*} 


\begin{table}[b]
\caption{\label{tab:table1} Lattice parameters and linear expansivities ($10^{-6}$/\,K) measured by experiment and calculated by MD simulation}
\begin{ruledtabular}
\begin{tabular}{l  c  c  c  c c  }
 & $a\,(\AA)$ & $\alpha_a$& $c$\,(\AA)&$\alpha_c$ & $u$  \\\hline
Exp(300\,K) & 4.592 & & 2.958 & & 0.3092  \\
Exp(1373\,K)& 4.635 &8.73 & 2.993 &11.0 & 0.3123 \\
Cal(300\,K) & 4.565 & &2.935 & & 0.3058 \\  
Cal(1373\,K)& 4.603 &8.75 & 2.967 &10.2 &0.3087   
\end{tabular}
\end{ruledtabular}
\end{table}

\begin{table*}[t]
\caption{\label{tab:table2} Frequencies (meV) of the transverse acoustic modes of rutile at ambient and high temperature, from experiment and from calculations with QHA, MD and TDEP methods}
\begin{ruledtabular}
\begin{tabular}{l  c  c  c  c  c  c  c  c  c}

$ $ & X & $(0.5,0,0.25)$&R&Z&$(0,0,0.25)$&$\Gamma$&M&$(0.5,0.5,0.25)$&A\\ \hline
&&&& $T$ = 300\,K &&&&&\\\hline
Exp  \footnote{Experimental data from Ref. [\onlinecite{Neutron1971}]}& 12.77 &       &       & 15.25 &      &  0  & 12.15 &      & \\
QHA \footnote{First principles calculations from Ref. [\onlinecite{Sikora2005}]}  & 11.04 & 17.30 & 10.42 & 10.42 & 7.15 &  0  & 9.55  & 5.65 & 9.92\\
QHA \footnote{First principles calculations from Ref. [\onlinecite{Mitev2010}]} & 13.52 &       & 13.02 & 12.90 &      &  0  & 9.05  &      & 12.03\\
QHA & 11.68 & 17.67 & 12.30 & 12.60 & 8.33 &  0  & 9.15  & 7.09 & 12.09\\
MD  & 13.60 & 19.88 & 13.49 & 15.36 & 10.36&  0  & 10.95 & 9.90 & 13.07\\
TDEP& 13.60 & 22.27 & 8.51  & 16.26 & 10.40&  0  & 11.01 & 6.01 & 9.20\\\hline
&&&& $T$ = 1373\,K &&&&&\\\hline
QHA \footnote{First principles calculations from Ref. [\onlinecite{Mitev2010}] with ambient temperature but 1\% isotropic lattice expansion} & 5.81  & 12.85 & 4.09  & 7.61  & 3.06\,i& ($>$7)\,i& 4.28  & 5.75\,i& 8.63\\
QHA & 3.13  & 12.59 & 2.02  & 9.07  & 3.78\,i& 9.83\,i&3.50  & 5.82\,i& 9.83\\
MD  & 15.41 & 19.81 & 16.93 & 18.04 & 13.34&  0  & 12.27 & 13.78 &14.28\\
TDEP& 16.91 & 21.12 & 13.53 & 19.63 & 13.35&  0  & 11.07 & 11.48 &11.57
  
\end{tabular}
\end{ruledtabular}
\end{table*}

\section{Calculations}

\subsection{First principles molecular dynamics simulations and techniques for calculating the vibrational spectra}

First-principles calculations using the local density approximation (LDA) of  density functional theory (DFT) \cite{lda} were performed with the VASP package.\cite{vaspa, vaspc}
Projector augmented wave pseudopotentials and a plane wave basis set with an energy cutoff of 500 eV were used in all calculations. 
Previous work showed that for best accuracy, the Ti pseudopotential should treat the semicore 3$s$ and 3$p$ states as valence,\cite{Harrison2004,  Mitev2010, Refson2013} 
and we took this approach with a similar LDA functional. 
Our calculated  elastic properties, lattice dynamics, and dielectric properties derived from the optimized structure for 0\,K, were in good agreement with results from experiment and from previous DFT calculations.
Phonon dispersions and phonon DOS spectra were obtained in the quasiharmonic approximation
by minimizing the vibrational free energy as a function of volume
\begin{eqnarray}
\! \! \! \! \! \! F(V,T) = E_0 + \int\limits_{- \infty}^{+ \infty} g( \omega ) \left(  \frac{\hbar \omega}{2}  + k_{\rm B }T \ln(1 - {\rm e}^{- \hbar \omega / k_{\rm B }T})  \right)
{\rm d} \omega \; ,
\end{eqnarray}
where $E_0 $ is the energy calculated for the relaxed structure at $T = 0 \,$K.

First-principles Born-Oppenheimer AIMD simulations for a $2 \times 2 \times 4$ supercell and a $2 \times 2 \times 1$ $k$-point sampling were performed to thermally excite phonons to the target temperatures of 300 and 1373\,K. 
For each temperature, the system was first equilibrated for 3\,ps as an $NVT$ ensemble with temperature control by a Nos{\'{e}} thermostat, then simulated as an $NVE$ ensemble for 20\,ps with time steps of 1\,fs. 
Good relaxations with residual pressures below 0.5\,GPa were achieved in each calculation that accounted for  thermal expansion. The lattice parameters and thermal expansivities determined by the simulations are presented in Table \ref{tab:table1}.  

\subsubsection{Fourier transformed velocity autocorrelation method}

Velocity trajectories were extracted from the MD simulation at each temperature, and were then transformed to the corresponding vibrational energy and/or momentum domain.\cite{Koker2009, MD2010, Lan2012, Lan2014, thesis}
Because the FTVAC method does not assume a form for the Hamiltonian,
it is a robust tool for obtaining  vibrational spectra from MD simulations, even with strong anharmonicity.
The phonon DOS is given by
\begin{equation}
\label{eq:DOScal}
g(\omegaup)=\sum_{n,b} \,\int e^{-{\rm i} \omegaup t}  \langle \vec{v}_{n,b}(t)\,\vec{v}_{0,0}(0) \rangle \, {\rm d} t \; ,
\end{equation}
where $\langle \, \rangle$ is an ensemble average, and $\vec{v}_{n,b}(t)$ is the velocity of the atom $b$ in the unit cell $n$ at time $t$.
Further projection of the phonon modes onto each $k$ point in the Brillouin zone was performed by computing the phonon power spectrum with the FTVAC method, with a resolution determined by the size of the supercell in the simulation.  

\subsubsection{Temperature-dependent effective potential method}

In general, 
the cubic phonon anharmonicity contributes  
to both the phonon energy shift and the 
lifetime broadening, whereas 
the quartic anharmonicity contributes only to the phonon energy shift.\cite{Maradudin, Lan20122}
To distinguish the roles of cubic and quartic anharmonicity, 
the temperature-dependent effective potential (TDEP) method \cite{Hellman1, Hellman3} was used.  
In the TDEP method, 
an effective model Hamiltonian is used to sample the potential energy surface,
not at the equilibrium positions of atoms,
but at the most probable positions for a given temperature in an MD simulation \cite{Hellman1}
\begin{eqnarray}
\label{eq:hamil}
H &=& U_0+\frac{1}{2}\sum_i m {\bf p}_i^2+\frac{1}{2}\sum_{ij\alpha\beta} \phi_{ij}^{\alpha\beta}u_i^{\alpha} u_j^{\beta}\\ \nonumber
&&+\frac{1}{3!}\sum_{ijk\alpha\beta\gamma} \psi_{ijk}^{\alpha\beta\gamma}u_i^{\alpha} u_j^{\beta} u_k^{\gamma} \; ,
\end{eqnarray}
where $\phi_{ij}$ and $\psi_{ijk}$ are  second- and third-order force constants, ${\bf p}$ is  momentum,
and $u_i^\alpha$ is the Cartesian component $\alpha$ of the displacement of atom $i$.  
 %
In the fitting, the ``effective'' harmonic force constants $\phi_{ij}$ are renormalized by
the quartic anharmonicity. 
The cubic anharmonicity, however, is largely accounted for by the third-order force constants $\psi_{ijk}$, 
and can be understood  in terms of the third-order phonon self-energy  
that causes linewidth broadening.\cite{Maradudin, Chen2014} 

The resulting Hamiltonian
was used to obtain the renormalized phonon dispersions (TDEP spectra) accounting for both the anharmonic shifts $\Delta$, and broadenings  $\Gamma$, 
of the mode $\vec{q}j$. These are derived from the real and imaginary 
parts of the cubic self-energies $\Sigma^{(3)}$, respectively.\cite{Maradudin}

\begin{eqnarray}
\Delta(\vec{q} j;\Omega)&=& -\frac{18}{\hbar^2}\sum_{{ \vec{q}_1}j_1} \sum_{{\vec{q}_2}j_2} \big\vert V(\vec{q}j;{ \vec{q}_1}j_1;{ \vec{q}_2}j_2) \big\vert^2 \Delta(\vec{q}_1+\vec{q}_2-\vec{q})\nonumber \\
  &&\times \,  {\wp} \Big[ \frac{n_1+n_2+1}{\Omega+\omegaup_1+\omegaup_2}
-\frac{n_1+n_2+1}{\Omega-\omegaup_1-\omegaup_2} \nonumber \\
  &&\mbox{} \quad +\frac{n_1-n_2}{\Omega-\omegaup_1+\omegaup_2}
-\frac{n_1-n_2}{\Omega+\omegaup_1-\omegaup_2} \Big] \label{shift3}\\
\Gamma(\vec{q}j;\Omega)&=& \frac{18\pi}{\hbar^2}\sum_{{ \vec{q}_1}j_1} \sum_{{\vec{q}_2}j_2} \big\vert V(\vec{q}j;{\vec{q}_1}j_1;{\vec{q}_2}j_2) \big\vert ^2 \Delta(\vec{q}_1+\vec{q}_2-\vec{q})\nonumber   \\
 &&\times \big[ (n_1+n_2+1) \, \delta(\Omega-\omegaup_1-\omegaup_2)\nonumber \\
&&\mbox{} \quad +2(n_1-n_2) \, \delta(\Omega+\omegaup_1-\omegaup_2) \big]    \; ,
\label{broadening3}
\end{eqnarray}
where $\Omega$ is the renormalized phonon frequency and $\wp$ denotes the Cauchy principal part. 
The $V(.)$'s are elements of the Fourier transformed third order force constants $\psi_{ijk}$ obtained in the TDEP method. The $\Delta(\vec{q}_1+\vec{q}_2-\vec{q})$ ensures conservation of momentum.

\subsection{Crystal orbital Hamilton population method}

The Crystal Orbital Hamilton Population Method (COHP)
projects plane waves to local orbital basis functions and is used to partition the band-structure energy into the bonding, nonbonding and antibonding
interactions between two atoms. COHP provides a semiquantitative interpretation of the net bonding characteristics.

In this work, COHP spectra were calculated with the package
Local-Orbital Basis Suite Towards Electronic-Structure
Reconstruction (LOBSTER).\cite{cohp, cohp2}
The phonon displacement patterns were generated in a $2 \times 2 \times 4$ supercell with a $8 \times 8 \times 4$ $k$-point sampling according to the eigenvectors of the mode. Self-consistent wavefunctions were then computed
and projected onto a local basis of Slater type orbitals with proper orthonormalization, giving the COHP spectra of the system.  Since
negative values of the COHP suggest bonding contributions, --\,COHP spectra are presented, following convention.

\subsection{Results}

As shown in Fig.~\ref{fig:DOS}(b), 
the neutron-weighted
DOS spectra calculated with FTVAC method reproduced the thermal shifts and broadenings  
observed in the experimental DOS of Fig.~\ref{fig:DOS}(a).  
In particular, upon heating to 1373\,K, an anomalous stiffening of peak 1 by 2.1\,meV was obtained. 

Figure~\ref{fig:quartic} shows the 
vibrational energies of  the TA branch,
calculated by the FTVAC method with AIMD trajectories. 
From 300 to 1373 \,K, the TA branch increases in energy by an average 
of about 2.1\,meV. 
Especially for this TA branch, 
Fig.~\ref{fig:quartic}(b) shows an enormous 
discrepancy of phonon energies between the MD calculation and the QHA (orange dashed line) at 1373\,K. 
The unstable phonon modes predicted by the QHA  are fully stable   
in the AIMD simulations at high temperatures, however.

Using the same MD trajectories as for the FTVAC method, 
the calculated TDEP dispersions agree well with the FTVAC results as shown in Fig.~\ref{fig:quartic}(a)(b), and also in Table \ref{tab:table2}. 
(The full TDEP dispersions are shown in the Supplemental Material.\cite{supplemental}) 
The cubic anharmonicity of rutile TiO$_2$ 
is strong for phonons at energies above 25\,meV, \cite{Lan2012}
causing broadening of the phonon DOS
of Fig. \ref{fig:DOS} and broadening of the dispersions.\cite{supplemental}
Nevertheless, at 1373\,K the TA modes below 20\,meV have 
only small linewidth broadenings.
Furthermore, they  are  close in energy to those calculated 
if all $\psi_{ijk}$ are set to zero in 
Eq. \ref{eq:hamil}, showing the dominance of quartic anharmonicity and the
small cubic anharmonicity of the TA modes.

\begin{figure*}[t]
\includegraphics[width=1.9\columnwidth]{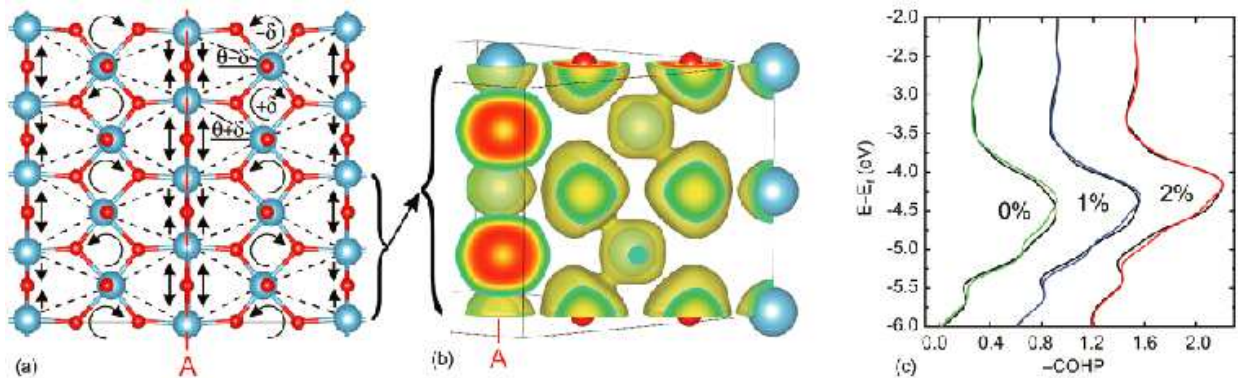}
\caption{(a) Displacements of atoms for the TA mode at the R-point in the (1-10) plane. Light blue spheres are Ti atoms, and O atoms are red. Arrows depict distortions of the structural units (dashed rhombuses). The rotational movements of structures, or the ``ring" patterns, are indicated with circled arrows. 
(b) ELF isosurface of a ``ring" shown in (a) with the value of 0.3. The ELF increase is apparent in the bond of shorter distance owing to the ring displacement. 
The ELF is the probability measure of finding an electron at a
location given the existence of neighbouring electrons.
It ranges from 0 (no electron) to 1 (perfect localization).  
(c) COHP analysis of Ti--O bonds for equilibrium lattice parameter at $T$=0\,K (0\%), and for linear expansions of 1\% and 2\%. Shown in color are --COHP results for the same structures with the phonon of panel (a) having 0.14\,${\rm \AA}$ normal displacements of Ti atoms. 
Curves for 1\% and 2\% expansion are offset  by 0.6 and 1.2. }
\label{fig:pattern}
\end{figure*}

\section{Discussion}

\subsection{Phonon quarticity of TA branch}

For more details about  the anomalous anharmonicity of the TA modes, 
the frozen phonon method was used to calculate potential energy surfaces for specific phonons, as in
Fig.~\ref{fig:quartic}(c). 
At 300\,K the potential energy of the TA mode at the $R$-point is nearly quadratic,
with a small quartic part. 
With the lattice expansion characteristic of 1373\,K, 
the potential energy curve transforms to being nearly quartic. 

In fact, for all modes in the TA branch that were evaluated by the frozen phonon method, 
the potential energy surface develops a quartic form with lattice expansion
(see Supplemental Material \cite{supplemental}).
Even more interesting behavior occurs  along the directions of $M$--$A$, $\Gamma$--$Z$ and $\Gamma$--$X$, 
where the  potential energy surface develops a Landau-type 
form, with negative curvature at zero displacement,
but a positive 
quartic shape at large displacements.
(The height of the energy barrier between the two minima is  lower than $k_{\rm B}T$,
however.)
For a quantum quartic oscillator, the vibrational frequency stiffens with temperature
owing to the increasing spread between the energy levels.\cite{Dorey1999,ScF3}
We  assessed a high temperature behavior 
by assigning
Boltzmann factors to the different oscillator
levels derived from frozen phonon potentials, giving 
the energies of the quartic TA modes at 1373\,K. 
As shown in Fig.~\ref{fig:quartic}(b),  they are reasonably close to the FTVAC and TDEP results.


\subsection{Phonon-tracked hybridization and its change with thermal expansion}

The  patterns of atom displacements 
in the anomalous modes at $\Gamma$, R, and along Z-$\Gamma$, $\Gamma$-M, and M-A
were identified, and  those for the R point in Fig. \ref{fig:pattern}(a) are typical.
In these anomalous modes, the O atoms were approximately stationary,
and each O atom has a Ti neighbor that moves towards it and another Ti neighbor that moves away from it by approximately the same amount.
These modes have ``ring'' patterns in which displacements of Ti atoms rotate  
a structural unit, and all the O atoms see approximately the same change in their Ti neighborhood.
In the positive and negative displacements of these modes,
the O atoms show
an accumulation of charge in the Ti--O bond of shorter distance
and a depletion in the bond of longer distance, as indicated by a much higher value of the electron localization function (ELF) \cite{ELF} for the short bond 
shown in Fig.~\ref{fig:pattern}(b).
%


We calculated the ``bond-weighted'' electron DOS by partitioning the band structure energy into bonding and no-bonding contributions and obtaining the crystal orbital Hamilton population (COHP) spectrum \cite{cohp, cohp2} of rutile with different lattice parameters in a $2 \times 2 \times 4$ supercell.  Figure \ref{fig:pattern}(c) shows the COHP spectrum of the 
bonds formed by the Ti\,$3d$\, and O\,$2p$ orbitals between
5.5 and 3.5\,eV below the Fermi energy. 
With lattice expansion (of 1\% or 2\%), these bonding states become less favorable, and their COHP spread
narrows. 
Also shown in color in Fig. \ref{fig:pattern}(b) is the COHP with a frozen phonon mode at the R-point
having 0.14\,${\rm \AA}$ displacements of  Ti atoms. 
On the scale of thermal energies, the broadening effect from the  phonon changes considerably with lattice expansion. 
For the equilibrium lattice parameter (labeled ``0\%''), 
bonding states at both the top and bottom of the Ti\,$3d$\, -- O\,$2p$ 
band are shifted upwards by the phonon.
With larger lattice parameter, however, the changes take place mostly in the 
states at the bottom of the band, which broadens the COHP spectra and helps to lower the bonding energy. Additional bonding states also appear near --3 eV.
The overall effect is that the potential energy is relatively insensitive to small displacements,
and the phonon potential has a much flatter bottom with 1\% lattice expansion as shown in Fig.~\ref{fig:quartic}(c). 
For a 2\% linear expansion, a Landau-type phonon potential evolves, with local minima away from 
zero displacement owing to the spread of states
to the lowest part of the band. 

\begin{figure}[t]
\includegraphics[width=0.7\columnwidth]{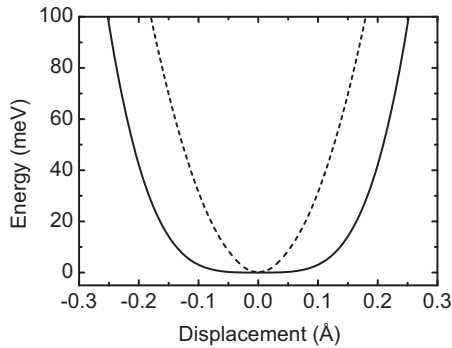}
\caption{Phonon potentials for the ``ring" pattern of Fig. \ref{fig:pattern}(a). The solid curve is the potential of the R-point mode at 1373\,K as in Fig.~\ref{fig:quartic}(c). The dashed curve is the potential of the same mode 1373\,K, but with the  ``ring" pattern broken by immobilizing the Ti atoms along the A-direction in Fig. \ref{fig:pattern}(a). The shape of dashed curve is quadratic. }
\label{fig:ringfreeze}
\end{figure}

The effect can be understood by adapting a hard sphere model for
the chain of -Ti-O-Ti-O- atoms along the direction ``A'' or around the ``ring'' units in the 
phonon displacement pattern in Fig. \ref{fig:pattern}(a).
With an increase in lattice parameter,
the longer Ti--O bond makes a smaller contribution to the interatomic force during its vibrational cycle.
The shorter Ti--O bond gives a stronger hybridization of  Ti\,$3d$\, -- O\,$2p$ orbitals as the Ti atom moves closer to the O atom.
The hybridization serves to offset the energy of short-range repulsion. 
With lattice expansion, the short-range repulsion is weaker, and 
hybridization favors electrons  
between the shorter Ti--O pairs in the phonon displacement pattern. 
The ``ring'' patterns of the phonons play an important role in increasing the degree of hybridization as they 
complete the electron back-donation cycles from the O to the Ti atoms. For example, a 3\% decrease of Bader effective charge (+2.22 at equilibrium) was found for the Ti atoms with 0.14 \,${\rm \AA}$ displacments in the ``ring" patterns, which is comparable to the charge decrease of the Ti atoms during the ferroelectric transition of rutile.\cite{Harrison2004} However,
if Ti atoms along the direction ``A'' were locked down at their 
equilibrium positions so the ring motion is broken, the resulting decrease of the effective charge dropped  by 50\% to 70\% in the ``ring" patterns. The potential 
was found to rise, and was largely quadratic even at 1\% or 2\% expansion, as shown in Fig.~\ref{fig:ringfreeze}.

A macroscopic elastic response to this phonon can also be identified with the assistance of Fig. \ref{fig:pattern}(a). 
In equilibrium, the apex angles of the rhombuses are all equal, but with the rotation by $\delta$, the vertical stretching of  rhombuses along the line A is $ 2a \sin ( \theta + \delta )$, and the contraction is $ 2a \sin ( \theta - \delta )$, where $\theta$ is the semi-angle of the rhombus. For small $\delta$, a Taylor expansion gives a net vertical (or horizontal) distortion of $  - 2a \delta^2 \sin \theta $ (or $  - 2a \delta^2 \cos \theta$). The distortions are proportional to $\delta ^2$, while the atom displacements in this TA mode are proportional to $\delta $. 
A strain energy that goes as the square of this distortion is consistent with a quartic potential. 
 
The hybridization in the Ti--O bond  is very sensitive to interatomic distance, much as has been noticed 
in the ferroelectric distortion of BaTiO$_3$.\cite{ferro1}
For rutile TiO$_2$, however, the 
hybridization follows the atom displacements in thermal phonons (instead of a displacive phase transition),
and  this  ``phonon-tracked hybridization''  changes with lattice parameter. 
It provides a source of extreme phonon anharmonicity, but also provides thermodynamic stability for  rutile TiO$_2$.
It may occur in other transition metal oxides that show unusual changes of properties with lattice parameter or with structure, and such materials may be tunable with composition or pressure to control this effect. Besides altering thermodynamic phase stability, properties such as ferroelectricity and thermal transport will be affected directly. 

\section{Conclusion} 
Several inter-dependent methods proved useful for assessing the  anharmonic phonon dynamics of rutile TiO$_2$.
Ab initio methods of molecular dynamics, temperature-dependent effective potential, and frozen phonons
gave  accurate accounts of the thermal stiffening of 
the TA phonons that was measured by inelastic neutron scattering on  rutile TiO$_2$. 
These methods avoided the structural instability predicted by the quasiharmonic approximation (QHA).
With thermal expansion, many acoustic modes were unstable in the QHA,
as observed previously. 
From AIMD simulations and frozen phonon calculations, 
it was found that when the rutile structure undergoes thermal expansion, 
a hybridization between Ti and O atoms 
gives low-energy electron states when the Ti and O atoms are in closest proximity
during their vibrations. 
This phonon-tracked hybridization 
flattens the bottom of the  potential of the anomalous phonon modes,  
giving  quartic potentials that stabilize the rutile structure at high temperatures. 

\begin{acknowledgments}
Research at the SNS at the Oak Ridge National Laboratory was sponsored by the Scientific User Facilities Division, BES, DOE.
This work was supported by the DOE Office of Science, BES, under contract DE-FG02-03ER46055. 
\end{acknowledgments}

\bibliography{ref}


\pagebreak
\widetext
\begin{center}
\textbf{\large Supplemental Materials:\\
Phonon quarticity induced by changes in phonon-tracked hybridization during lattice expansion and its stabilization of rutile TiO$_2$}
\end{center}

\begin{figure*}[h]
\includegraphics[width=0.6\columnwidth]{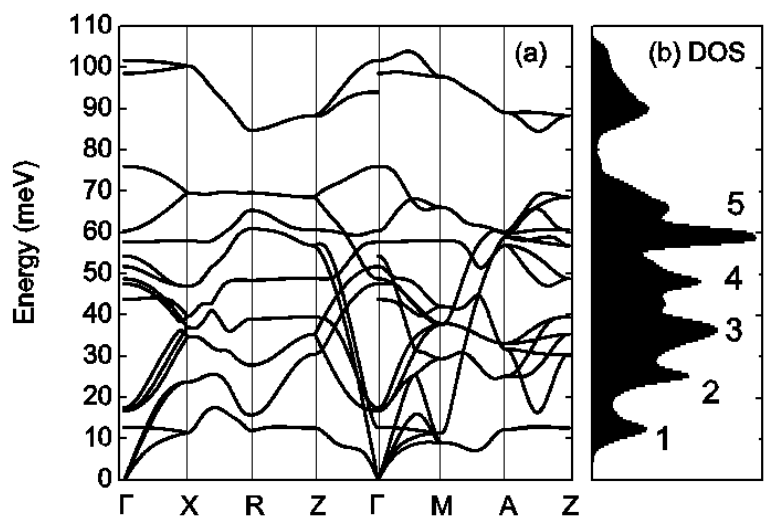}
\caption{(a) Calculated phonon dispersion of rutile TiO$_2$ along the high symmetry directions. (b) The corresponding phonon DOS. The DOS peaks are labeled by the same numbers as those in Fig.1 of the paper }
\label{fig:disp}
\end{figure*}

\begin{figure*}[b]
\includegraphics[width=1.0\columnwidth]{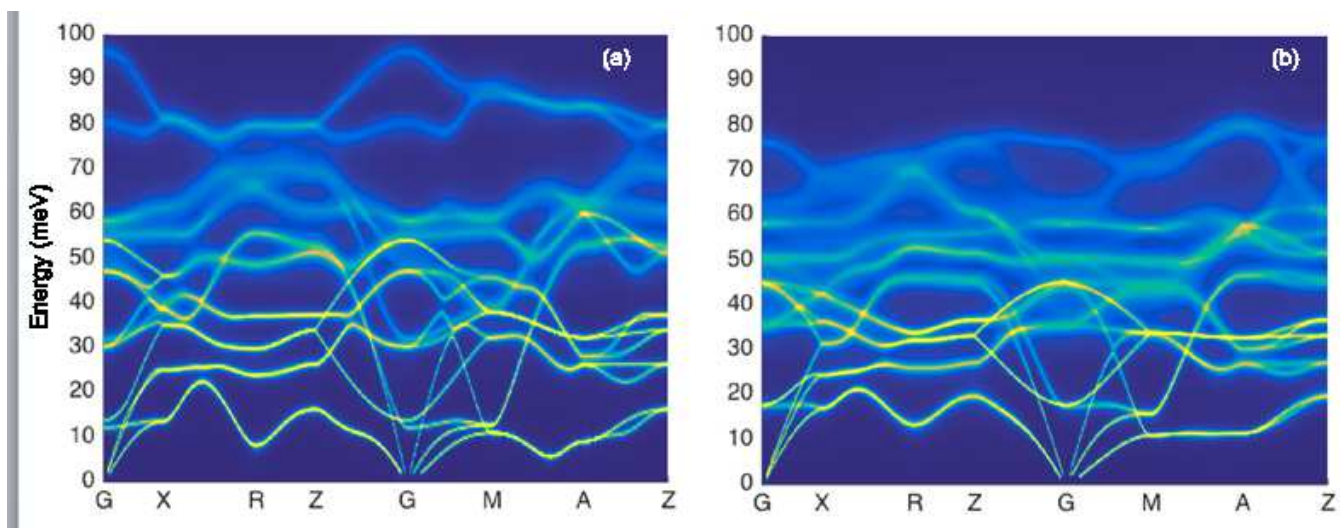}
\caption{TDEP phonon spectra along high symmetry points at (a) 300\,K and (b) 1373\,K}
\label{fig:fullTDEP}
\end{figure*}

\begin{figure*}[]
\includegraphics[width=1.0\columnwidth]{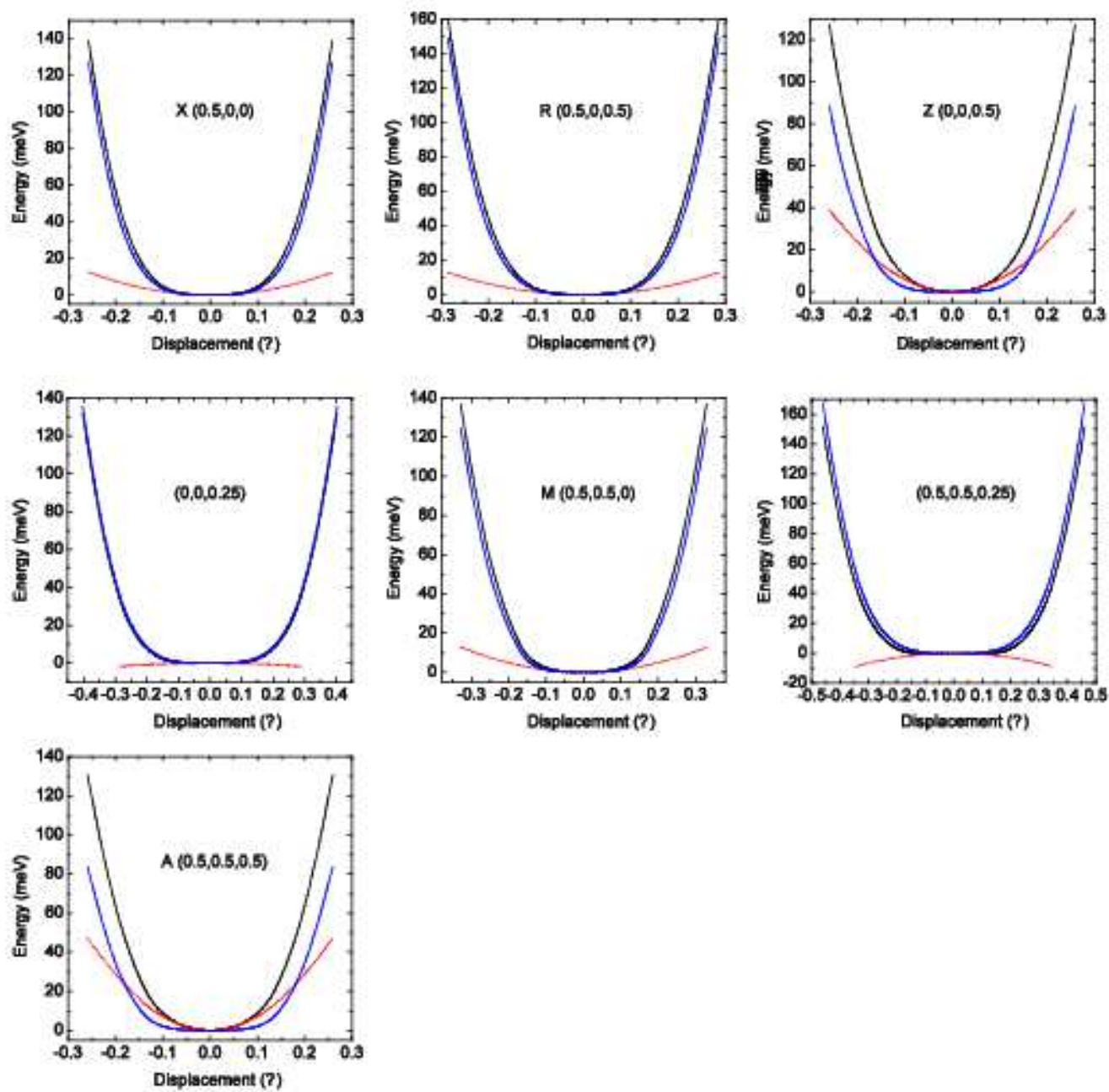}
\caption{Frozen phonon potentials (black) of TA modes at high symmetry points at 1373\,K. The potentials are decomposed into the harmonic component (red) and quartic component (blue). }
\label{fig:quarticlist}
\end{figure*}

 
\end{document}